\documentclass[%
pra%
,reprint%
,twocolumn%
,floatfix%
,superscriptaddress, showpacs%
,amssymb, amsmath, nobibnotes, aps]{revtex4-1}

\usepackage[latin1]{inputenc}
\usepackage{epsfig}
\usepackage{color}

\newcommand{\ind}[1]{_{\text{#1}}}
\newcommand{\bra}[1]{\left<#1\right|}
\newcommand{\ket}[1]{\left|#1\right>}

\begin{document}

\title{Finite-size effects in strongly interacting Rydberg gases}

\author{Martin G\"{a}rttner}
\affiliation{Max-Planck-Institut f\"{u}r Kernphysik, Saupfercheckweg 1, 69117 Heidelberg, Deutschland}
\affiliation{Institut f\"{u}r Theoretische Physik, Ruprecht-Karls-Universit\"{a}t Heidelberg, Philosophenweg 16, 69120 Heidelberg, Germany}
\affiliation{ExtreMe Matter Institute EMMI,
              GSI Helmholtzzentrum f\"ur Schwerionenforschung GmbH,
              Planckstra\ss e~1,
              64291~Darmstadt, Germany}

\author{Kilian P.~Heeg}
\affiliation{Max-Planck-Institut f\"{u}r Kernphysik, Saupfercheckweg 1, 69117 Heidelberg, Deutschland}

\author{Thomas Gasenzer}
\affiliation{Institut f\"{u}r Theoretische Physik, Ruprecht-Karls-Universit\"{a}t Heidelberg, Philosophenweg 16, 69120 Heidelberg, Germany}
\affiliation{ExtreMe Matter Institute EMMI,
              GSI Helmholtzzentrum f\"ur Schwerionenforschung GmbH,
              Planckstra\ss e~1,
              64291~Darmstadt, Germany}

\author{J\"{o}rg Evers}
\affiliation{Max-Planck-Institut f\"{u}r Kernphysik, Saupfercheckweg 1, 69117 Heidelberg, Deutschland}


\begin{abstract}
The scaling of the number of Rydberg excitations in a laser-driven cloud of atoms with the interaction strength is found to be affected by the finite size of the system. The scaling predicted by a theoretical model is compared with results extracted from a numerical many-body simulation. We find that the  numerically obtained scaling exponent in general does not agree with the analytical prediction. By individually testing the assumptions leading to the theoretical prediction using the results from the numerical analysis, we identify the origin of the deviations, and explain it  as arising from the finite size of the system. Furthermore, finite-size effects in the pair correlation function $g^{(2)}$ are predicted. Finally, in larger ensembles, we find that the theoretical predictions and the numerical results agree, provided that the system is sufficiently homogeneous.
\end{abstract}

\maketitle

%
\section{Introduction}

During the past few years, a growing amount of research, both in experiment and theory, has been aiming at a detailed understanding of ultracold gases of atoms excited to states of high principal quantum number. Such Rydberg atoms have extreme properties, of which most important are their long-range dipolar interactions. These imply the so-called dipole-blockade predicted in \cite{lukin2001, jaksch2000}, a suppression of further Rydberg excitation in the vicinity of one such excited atom, which is caused by an interaction-induced level shift. First experimental detection of the dipole blockade was reported in Refs.~\cite{tong2004} and \cite{singer2004}. Subsequently, further effects such as collective oscillations and superradiance, to name a few, were found \cite{comparat2010}. At sufficiently low temperatures, the atomic motion and collisions can be neglected on the timescale of the electronic dynamics, allowing for the `frozen-gas' approximation. These properties, combined with high control over the atomic positions provided by optical trapping techniques, make up an ideal model system for strongly correlated ensembles of atoms. Various applications in quantum information processing \cite{saffman2010} and quantum simulation \cite{weimer2010} are being pursued. With recent progress, also spatially resolved measurements  are coming within reach~\cite{schwarzkopf2011, guenter2012, olmos2011}.

As a particular consequence of the many-body correlations, the ground state of a strongly interacting and laser-driven Rydberg ensemble can undergo a second-order quantum phase transition~\cite{weimer2008}. At the critical point, determined by resonant laser driving, the system can be characterized by a single parameter $\alpha$ which depends on the laser Rabi frequency $\Omega$, the ensemble density and the interaction strength. The number of excited Rydberg atoms $N\ind{Ryd}$ was predicted to scale as $N\ind{Ryd}\sim \alpha ^ \nu$, with an exponent $\nu$  in the critical region  $\alpha \ll 1$. This result was found to agree well with numerical results from mean-field calculations and many-body simulations~\cite{weimer2008}.

In this paper, we revisit this scaling of the amount of Rydberg excitation with the interaction strength and show that it can be modified by finite-size effects in smaller ensembles of Rydberg atoms. For this, we compare the scaling prediction obtained from a theoretical model with results extracted from a numerical many-body simulation. We find algebraic scaling in both the model and the simulation. But the  numerically obtained scaling exponent $\nu$ in general does not agree with the analytical prediction. By testing the assumptions entering the argument leading to the prediction we identify the deviations to arise from the finite size of the system. To be precise, the estimation of the blockade radius based on the comparison of the two energy scales present in the system turns out not to be biased by the finite size of the system. On the other hand, relating the size of the blockade spheres to the steady-state excited fraction is found to be non-trivial in finite-size systems. In this context we discuss the validity of the so called 'super-atom' picture in finite systems. We also find finite-size effects in the pair correlation function $g^{(2)}$.
We finally analyze larger ensembles, imposing periodic boundary conditions. We find that, provided that the ensemble is sufficiently homogeneous, finite-size effects disappear and the numerically obtained scaling exponent agrees well with the predicted one. 
We point out that, while finite-size effects appear in the scaling exponent of the bulk number of Rydberg atoms, their interpretation crucially relies on a spatially resolved analysis of the excitation dynamics. 

Our study is not least motivated by the fact that a number of recent experiments approach length scales for which finite size effects are expected to become relevant~\cite{ditzhuijzen2008, tauschinsky2010, leung2011, viteau2011, kuebler2010}. In particular, the system sizes considered in the following are well within reach of experimental realizations of quasi 1D Rydberg ensembles~\cite{viteau2011}.

After introducing the model and numerical methods in Sect.~\ref{sec:model} we analyze, in Sect.~\ref{sec:dynamics},  the dynamical evolution of the ensemble under the influence of the laser radiation. In Sect.~\ref{sec:bulk}, we study bulk properties of the ensemble and show that the numerical simulations predict scaling parameters different from the analytical model. To identify the origin of this deviation, we compute spatially resolved observables in Sect.~\ref{sec:corr}. In Sect.~\ref{sec:pBC} we compare to systems with periodic boundary conditions. Section~\ref{sec:g2} deals with finite-size effects in the pair correlation function. We summarize our results in Sect.~\ref{sec:conclusion}.

%
\section{Model and methods}
\label{sec:model}

%
\subsection{Hamiltonian}

Our model system is a cloud of two-level atoms consisting of a ground state $|g\rangle$ and a Rydberg state $|r\rangle$. The atoms reside at fixed positions throughout the simulation time. In most experiments, the atoms are excited from $|g\rangle$ to $|r\rangle$ via a two-photon transition coupling to an intermediate level $|m\rangle$. However, if the intermediate level is far detuned it can be adiabatically eliminated \cite{honer2010}. In the present work we assume that the intensity and wavelength of the laser is temporally and spatially constant.

The Hamiltonian describing such a cloud of $N$ interacting, laser-driven two-level atoms can be written as~\cite{saffman2010,comparat2010,younge2009}
\begin{equation}
\label{eq:H1}
 H=H\ind{1B}+H\ind{int}.
\end{equation}
Here, $H\ind{1B}$ is the one-body Hamiltonian of $N$ non-interacting atoms, written in rotating wave approximation and in units where $\hbar=1$ as
\begin{subequations}
\begin{align}
 H\ind{1B} &= H_{\Delta} + H_{\Omega}\,,\\
 H_{\Delta} &= -\Delta \sum_{i=1}^N s^{(i)}_{rr}\,,\\
 H_{\Omega} &=  \frac{\Omega}{2}\sum_{i=1}^N  \left(s^{(i)}_{rg} + s^{(i)}_{gr} \right)\,,
\end{align}
\end{subequations}
where we used the notation $s^{(i)}_{\alpha\beta}=\ket{\alpha}_i\bra{\beta}$ for the internal projector of the $i$th atom.
$H_{\Delta}$ depends on the detuning $\Delta=\Delta_1+\Delta_2 = \omega_{l1}+\omega_{l2}-\omega_{gm}-\omega_{mr}$ between the laser frequencies $\omega_{li}$ and the energy differences $\omega_{\alpha\beta}=E_{\beta}-E_{\alpha}$ between the ground and intermediate, and the intermediate and Rydberg states. $H_{\Omega}$ contains the two-photon Rabi-frequency $\Omega$ that depends on the laser intensities and the dipole matrix elements of the transitions. We have assumed $\Omega$ to be real.

The long-range interactions between two Rydberg atoms are accounted for by the van der Waals term
\begin{equation}
 H\ind{int} 
 = \sum_{i<j}^N 
\frac{C_6}{|\mathbf r_i - \mathbf r_j|^6}
s^{(i)}_{rr}s^{(j)}_{rr}\,.
\end{equation}
%

In our simulations, we set $\Omega=1$, i.e., we measure energy in units of $\hbar\Omega$ and time in units of $1/\Omega$. 
We keep, however, $C_{6}$ as a dimensionful parameter and give lengths in micrometers, keeping in mind that increasing $C_6$ is equivalent to decreasing the volume at a fixed particle number, i.e., increasing the density. 

%
\subsection{\label{te}Time evolution}

We numerically solve the Schr\"{o}dinger equation
\begin{equation}
\label{eq:tdepSE}
 i\frac{\partial}{\partial t} \ket{\Psi} = H \ket{\Psi}
\end{equation}
for a set of $N$ two-level atoms at positions $\mathbf{r}_i$ and given parameters $C_6$
and $\Delta$, starting from an initial configuration where all atoms are in the ground state.
%
%
A variety of approaches has been suggested to deal with the large dimension of Hilbert space of the above system which scales exponentially as $d\sim2^N$.
These include techniques relying on an expansion in the order of correlations~\cite{tong2004, choita2008, weimer2008,schempp2010}, rate-equation methods \cite{ates2007a, ates2007b, ates2011,heeg2012}, and many-body simulations on truncated state spaces~\cite{robicheaux2005, hernandez2006, hernandez2008, pohl2009, pohl2010, younge2009, weimer2008, loew2009, olmos2009a, olmos2009b, tezak2011,gaerttner2012,komnik2012}. 
In the following analysis, we truncate the Hilbert space, making use of the fact that states $\ket{\Phi}$ with high diagonal elements $\bra{\Phi}H\ind{int} + H_\Delta\ket{\Phi}$ are less likely to become populated.
We note that for infinite interaction strength, starting from the ground state $\ket{\mathbf{0}}=\ket{gg\ldots g}$, only singly excited states are accessible. Higher excited states can be discarded because they have infinite interaction energy. The resulting dynamics is a Rabi oscillation between the ground state $\ket{\mathbf{0}}$ and the symmetrized singly excited state $\ket{\mathbf{s}}=N^{-1/2}\sum_{i=1}^N\ket{1;i}$ \cite{lukin2001}, with collective Rabi frequency $\Omega_c=\sqrt{N}\Omega$ \cite{hernandez2006}. Collective Rabi oscillations of two atoms could already be observed in experiment \cite{gaetan2009}. 

Taking into account the dipole-dipole interactions which inhomogeneously shift the transition frequency, two atoms cannot be excited simultaneously within the blockade volume $V_b$. All atoms in this volume share one excitation and their dynamics is governed by the collective Rabi-frequency $\sqrt{N_b}\Omega$. The blockade volume is found by comparing the relevant energies $C_6R_b^{-6}\approx \sqrt{N_b}\Omega$  where the blockade radius $R_b$ is the minimal distance that two simultaneously excited atoms can have. In a $d$-dimensional gas of density $\rho$ (atoms per unit length to the power $d$) we have $N_b=\rho V_b\propto \rho R_b^d$.
These considerations lead to~\cite{saffman2010,comparat2010,weimer2008}
\begin{equation}
\label{eq:Rb}
 R_b \propto \left(\frac{C_6}{\Omega\sqrt{\rho}}\right)^{{2}/{(d+12)}}\,.
\end{equation}
Note that in deriving $R_b$, we neglected the contribution of $H_\Delta$ to the diagonal element. The above estimate is therefore only valid near resonance, i.e., for $\Delta\ll\Omega$. 

We use Eq.~\eqref{eq:Rb} to truncate the Hilbert space by neglecting all states in which two excited atoms have a distance smaller than the cutoff distance $r_b<R_b$. This is achieved by choosing the proportionality constant in Eq.~\eqref{eq:Rb} appropriately. Other ways to reduce the number of basis states are to put an upper bound $m\ind{max}$ on the number of excitations or to exclude states with an interaction energy higher than some threshold $E\ind{int}^{\text{max}}$. We have checked that our numerical results are converged with respect to all truncation criteria, i.e., to the chosen values of the cutoff parameters.

To implement the truncation we construct, from the canonical set of product states $\ket{\alpha_1 \alpha_2\ldots \alpha_N}=\ket{\alpha_1}\otimes\ket{\alpha_2}\otimes\ldots\ket{\alpha_N}$ with $\alpha_i\in\{g,r\}$, 
basis states of the form $\ket{m;i_1,i_2\ldots i_m}$, i.e., states with $m$ atoms $i_1$, $i_2 \ldots i_m$ in the Rydberg state and all other atoms in the ground state. The detuning and interaction parts of the Hamiltonian are diagonal in this basis, while the Rabi terms connect states that differ in the internal level of one atom only. Therefore the Hamiltonian matrix can be divided into sub blocks consisting of states containing the same number $m$ of excited atoms. The Hamiltonian has a tridiagonal structure in these blocks since only states with $m$ differing by one are connected by the laser Hamiltonian. The blocks on the diagonal are diagonal in the above basis. This structure allows for a convenient truncation and construction of the resulting evolution matrices. After the truncation, we denote the $n$ remaining basis states of form $\ket{m;i_1,i_2\ldots i_m}$ as $\ket{\Phi_k}$ with $k\in\{1, \dots, n\}$.

While the performance of our code depends on the number of atoms it is crucially affected by the number of excitations that fit into the trap volume. The major numerical limitation is computation time. If we want to limit the time needed for the propagation of a single realization into the steady state to less than one hour on a state-of-the-art desktop processor (2.33~GHz), we are limited to state spaces of about four million states. Given this constraint, if we consider an atom number of about 100, the trap size needs to be limited such that it can contain at most six-fold excitations. For 60 atoms we are limited to about nine excitations. For such system sizes the coherent time evolution of the system does not fully converge to a steady state, with small oscillations persisting in the long-time limit. Moreover, for the calculation of spatially dependent observables such as the  distribution of the Rydberg excitations, a single run with about 100 atoms does not provide sufficient statistics. These problems can be overcome by artificially increasing the particle number by Monte Carlo sampling. This simply means that we calculate the time evolution of the system for several realizations (typically 1000 to 5000) with randomly chosen atom positions and average over the resulting observables. This method was used before in Refs.~\cite{younge2009, carroll2009, ryabtsev2010} and can be justified for Bose-Einstein condensates as well as for thermal clouds as explained in Ref.~\cite{vanbijnen2011}.

Our model exclusively contains the unitary time evolution, i.e., spontaneous emission, dephasing or laser linewidths are not explicitly accounted for. A proper treatment of such incoherent processes would require methods such as a master equation or a quantum Monte Carlo simulation of the incoherent jumps, which would considerably increase the computational complexity. From a physical point of view, this approximation can be justified if the experimental time scales are short compared to the life time of highly excited Rydberg states, and if stabilized laser systems with low linewidth are used.

%
\subsection{Observables}

For a single realization of our system, all information is contained in the coefficient vector $\mathbf{c}$ with elements $c_k$, 
\begin{equation}
 \ket{\Psi}=\sum_{k=1}^n c_k \ket{\Phi_k}\,,
\end{equation}
with basis states $\ket{\Phi_k}$ as defined in Sec.~\ref{te}. For each realization the observables are calculated from the state vector at a given time of interest and then averaged by Monte Carlo sampling.
The mean number of Rydberg atoms is
\begin{equation}
 N\ind{ryd} = \langle \hat{N}\ind{ryd} \rangle =  \sum_{k=1}^n |c_k|^2 m^{(k)}\,
\end{equation}
where $m^{(k)}$ is the number of excited atoms in state $\ket{\Phi_k}$.
The excitation probability of the $i$th atom is given by
\begin{equation}
 \langle \hat{N}^{(i)}\ind{ryd} \rangle = \sum_{k=1}^n |c_k|^2 \delta_{\alpha_i^{(k)} r}\,,
\end{equation}
where $\alpha_i^{(k)}$ is the internal state of atom $i$ in state $\ket{\Phi_k}$, and the Kronecker delta selects the coefficients of the states in which the $i$th atom is excited. 


The spatial distribution of the Rydberg excitations $N\ind{ryd}(\mathbf{r})$ is obtained by dividing the trap volume into small cells and summing up the excitation probabilities of all atoms that lie in such a cell, normalized  by  their number.
Most important in this work are spatial correlations quantified by the pair correlation function $g^{(2)}(\mathbf{r},\mathbf{r'})$~\cite{wuester2010}. This function is a measure for the conditioned probability of having an excitation at $\mathbf{r}$ if there is already one at $\mathbf{r'}$. Since the interaction potential in our model only depends on the mutual distance of two atoms, it is valid to assume that also $g^{(2)}$ only depends on $r=|\mathbf{r}-\mathbf{r'}|$. This has been verified numerically. The pair correlation of two particles is defined as~\cite{wuester2010}
\begin{equation}
 g^{(2)}_{i,j} = \frac{\langle s^{i}_{rr} s^{j}_{rr} \rangle}{\langle s^{i}_{rr} \rangle \langle s^{j}_{rr} \rangle}.
\end{equation}
Discretizing space, we define 
\begin{equation}
 g^{(2)}(r) = \frac{\sum_{i,j}^{(r,\Delta r)} g^{(2)}_{i,j}}
		   {\sum_{i,j}^{(r,\Delta r)} 1}
\label{eq:g2r}
\end{equation}
where $\sum_{i,j}^{(r,\Delta r)}$ denotes the sum over all pairs with mutual distance $|\mathbf{r}_{i} - \mathbf{r}_{j}|$ lying within the interval $[r,r+\Delta r]$. This means that we sum up the correlations of all these pairs divided by their number.
Note that some authors use the alternative definition $\tilde{g}^{(2)}=g^{(2)}-1$ \cite{guenter2012}. According to our definition $g^{(2)}(r)=1$ corresponds to uncorrelated atoms.

\begin{figure}[t]
 \centering
 \includegraphics[width=\columnwidth]{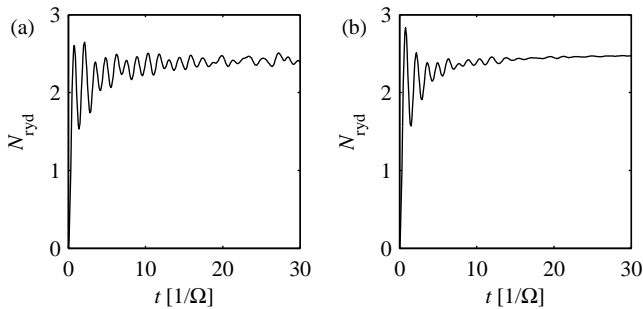}
 \caption{Time evolution of the number of Rydberg excitations in an ensemble of $N=50$ atoms enclosed in a cylindrical trap with radius $r=1\,\mu$m and length $L=10\,\mu$m. Further parameters are: $\Delta=0$, $C_6=900\,\Omega\,\mu$m$^6$. (a) Single realization.  The remaining oscillations at long times reflect the fluctuations of the positions of the Rydberg atoms. (b) Average over $1000$ Monte Carlo samples.}
 \label{fig:tdep}
\end{figure}

%
\section{Dynamical properties}
\label{sec:dynamics}

A single atom or a gas of non-interacting atoms undergoes simple Rabi oscillations on a time scale short compared to the dephasing time. If the laser is detuned, only a part of the population oscillates between the ground and excited states, while the rest of the population remains in the ground state. The oscillation frequency is modified to $\sqrt{\Delta^2+\Omega^2}$. In the case of infinitely strong interactions the system reduces to a two-level system showing collective Rabi oscillations between the ground state and the collective singly excited state as discussed in Sec.~\ref{sec:model}.

In the intermediate case, in which several atoms can be excited simultaneously, but interactions are non-zero, the dynamics is more complicated. In the example shown in Fig.~\ref{fig:tdep} we use a cylindrical trap of length $10\,\mu$m and diameter $2\,\mu$m. We observe that the Rydberg population of the cloud, initially being in the ground state, shows a saturation behavior. However, even for long times some oscillations remain, cf.\ Fig.~\ref{fig:tdep}(a). The strength of these oscillations depends on the number of excitations and on the number of atoms in the trap. This saturation can be interpreted in two ways: On the one hand, the mean-field picture predicts that, due to the disorder induced by the randomness of the atom positions, interaction shifts $\delta$ differ from atom to atom, leading to different Rabi frequencies $\sqrt{\delta^2 + \Omega^2}$ for the individual atoms. Consequently, the oscillations dephase over time and lead to a saturation of the overall excitation. Alternatively, one uses the so-called ``super-atom'' picture: Any excited atom in the ensemble blocks the excitation of the surrounding atoms which leads to collective excitation of the atoms within this blockade sphere. Again due to the random positions of the atoms, the number of atoms per blockade sphere and thus the collective Rabi frequency varies across the ensemble. The occurrence of different collective Rabi frequencies leads to a dephasing and thus to saturation.

Averaging over many realizations shows that the mean number of Rydberg excitations saturates fully after some time, cf.\ Fig.~\ref{fig:tdep}(b). The timescale of this saturation depends on how many Rydberg excitations are present in the gas, on the system size and also the detuning. Note that for very small traps that confine the gas to near one blockade volume, ongoing strong oscillations are observed even after Monte Carlo averaging. The reason is that in this case, only one or a few different collective Rabi frequencies are possible, depending on the number of atoms in the blockade spheres.
In the example shown in Fig.~\ref{fig:tdep}(b) we find that the excitation number converges to about $N\ind{ryd}=2.5$ Rydberg excitations. This means that $N_b=N/N\ind{ryd}=20$ atoms share one excitation and we expect to observe a collective Rabi-frequency of $\Omega_C/\Omega=\sqrt{N_b}=4.47$. Measuring the period of the initial oscillations in Fig.~\ref{fig:tdep}(b) we obtain $\Omega_C=2\pi/T=4.49\,\Omega$, consistent with the super-atom picture.

\begin{figure}[t]
 \centering
 \includegraphics[width=\columnwidth]{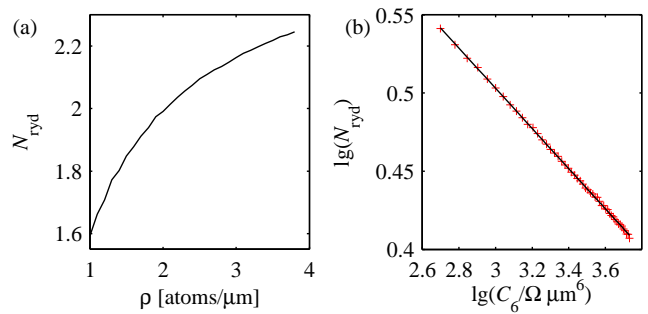}
 \caption{(Color online) Number $N_{\mathrm{Ryd}}$ of Rydberg excitations in a one dimensional gas. (a) $N_{\mathrm{Ryd}}$ as a function of the line density. Parameters: $L=10\,\mu$m, $C_6=900\,\Omega\,\mu$m$^6$, $\Delta=0$. (b) Number of excitations as a function of the van der Waals interaction constant, on a double logarithmic scale. Crosses: numerical simulation. Solid line: power-law fit $N_{\mathrm{Ryd}}\sim C_{6}^{-0.128}$. Parameters: $L=15\,\mu$m, $N=45$, $\Delta=0$ of $N=45$ atoms.}
 \label{fig:dens_C6_dep}
\end{figure}

%
\section{Finite-size effects}
\subsection{Signatures in bulk properties}
\label{sec:bulk}

In this section we study the properties of the system after is has saturated to its steady state.
For sufficiently high atomic density the number of Rydberg excitations does not scale linearly with the number of atoms, as it would be expected for a dilute non-interacting gas. Instead, a reduced Rydberg population is observed as is shown in Fig.~\ref{fig:dens_C6_dep}(a). This is caused by the well-known dipole blockade~\cite{tong2004} inhibiting the simultaneous excitation of two nearby atoms by means of a large interaction shift. This effect is the more pronounced the stronger the interactions are. Thus $N\ind{ryd}$ decreases as a function of the interaction parameter $C_6$, as is illustrated in Fig.~\ref{fig:dens_C6_dep}(b). Assuming that the blockade radius is given by Eq.~\eqref{eq:Rb}, one finds that~\cite{weimer2008}
\begin{equation}
\label{eq:scaling_exp}
 N\ind{ryd} \sim \frac{N}{N_b} \sim R_b^{-d} \sim C_6^{-{2d}/({d+12})}.
\end{equation}
From this, for a one-dimensional gas ($d=1$), one obtains an algebraic scaling of $N\ind{ryd}$ with $C_{6}$, with exponent $\nu=-2/13\approx -0.1538$. Fig.~\ref{fig:dens_C6_dep}(b) confirms that the excited fraction scales as a negative power of $C_6$, however, with a fitted exponent  $\nu\ind{fit}=-0.1283\pm0.0006$ (errors we give on fitted values just refer to the confidence interval of the fit). This is obtained for a one-dimensional trap of length $15\,\mu$m containing 45 atoms.
At first sight this result appears to be in contrast to the calculations reported in~\cite{weimer2008} where good agreement of numerical simulations with the analytically predicted value was reported. 
We will analyze this discrepancy in more detail in the following section and show that the deviation from the theoretical prediction is due to finite-size effects arising in our relatively small trap volume. 
For now we summarize that the excited fraction follows an algebraic scaling over a wide range of interaction strengths also in finite systems, while the scaling exponent deviates from the one predicted for an unbounded homogeneous gas.
%

\begin{figure}[t]
 \centering
 \includegraphics[width=\columnwidth]{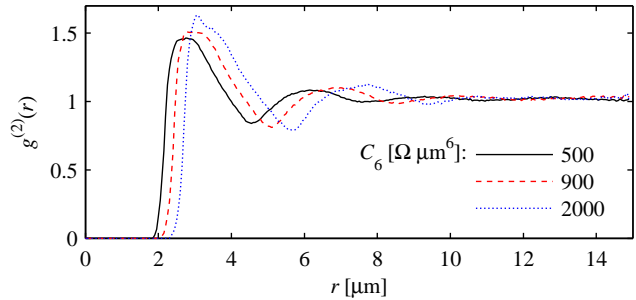}
 \caption{(Color online) Pair correlation function defined in Eq.~(\ref{eq:g2r}) for three different interaction strengths $C_{6}$ after $t=200\,\Omega$ evolution time. One-dimensional trap of length $L=15\,\mu$m holding $N=45$ atoms exposed to a resonant laser.}
 \label{fig:g2}
\end{figure}

\subsection{\label{sec:corr}Origin of the finite-size effects}

In the previous subsection we reported the numerically obtained power-law scaling of the number of Rydberg excitations $N\ind{ryd}\sim C_6^\nu$. The value for $\nu$ did not agree, however, with the value predicted and confirmed previously for a homogeneous gas. In the following, we trace back the origin of this discrepancy to finite-size effects. For this, we individually examine the two assumptions which lead to the prediction of the scaling parameter in Eq.~(\ref{eq:scaling_exp}). The first one is the estimate of the blockade radius, Eq.~(\ref{eq:Rb}). The second one is the assumptions that the number of Rydberg excited atoms can be obtained from an estimate based on the super-atom picture, $N\ind{ryd}\propto m\ind{max}=N/N_b=L/R_b$. Note that the effects studied here are physical finite-size effects in the sense that they originate from the finite trap volume rather than from computational limitations. Therefore, they could possibly be observed in experiments.

\subsubsection{Estimate for the blockade radius}

In order to analyze the estimate for the blockade radius given in Eq.~(\ref{eq:Rb}) we consider pair correlations. Fig.~\ref{fig:g2} shows the pair correlation function (\ref{eq:g2r}) of a one-dimensional gas in its thermalized state. Pairs with small mutual distance are never excited simultaneously, hence the $g^{(2)}$-function is zero for small $r$. Note that the numerical blockade radius ${r}_b$ was chosen well below the onset of non-zero correlations to ensure that the state space truncation does not affect the outcome of the simulations. The sharp peak which emerges just outside the blockaded region is located precisely at 
\begin{equation}
\label{eq:Rb_1D}
 R_b= \left(\frac{C_6}{\Omega \sqrt{N/L}}\right)^{2/13},
\end{equation}
which was our estimate of the blockade radius (Eq.~\eqref{eq:Rb}) leading to Eq.~(\ref{eq:scaling_exp}). 
\begin{figure}[t]
 \centering
 \includegraphics[width=\columnwidth]{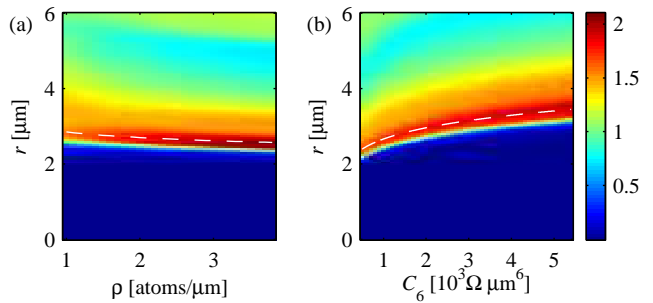}
 \caption{(Color online) Pair correlation function (\ref{eq:g2r}) as a function of (a) the atom line density, with $C_6=900\,\Omega\,\mu$m$^6$, $L=10\,\mu$m;  (b) the interaction constant $C_6$, with line density $\rho=3/\mu$m ($L=15\,\mu$m, $N=45$). Dashed lines indicate the estimates for $R_b$ as in Eq.~\eqref{eq:Rb_1D}.}
 \label{fig:g2_rho_C6}
\end{figure}
In order to probe this conjecture more thoroughly we analyze  $g^{(2)}$ for varying line density and interaction strength. Results are shown in  Fig.~\ref{fig:g2_rho_C6}. The predicted blockade radius perfectly matches the first maximum of $g^{(2)}$ in the whole parameter region. This indicates that the estimate for $R_b$ in Eq.~\eqref{eq:Rb} are reasonable even in finite geometries. We remark that in these figures one can also observe that correlations become stronger as the density or the interaction strength increases. We could also quantitatively verify Eq.~\eqref{eq:Rb_1D}. For this, we extracted the blockade radius from the simulation results shown in Fig.~\ref{fig:g2_rho_C6} by defining it as the pair distance at which the $g^{(2)}$-function first exceeds one. This observable shows a power law dependence over the whole range of parameters and fitting the exponent yields perfect agreement with Eq.~\eqref{eq:Rb_1D}.

\subsubsection{Estimate for the number of  Rydberg excited atoms}

\begin{figure}[t]
 \centering
 \includegraphics[width=\columnwidth]{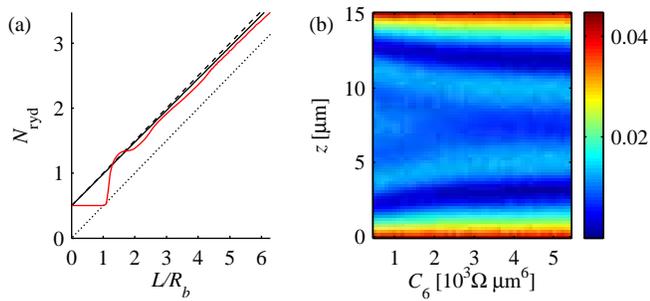}
 \caption{(Color online) Same setup as in Fig.~\ref{fig:g2}. (a) $N\ind{ryd}$ as a function of $L/R_b=L(C_6/\Omega\sqrt{\rho})^{-2/13}$. Red line: Numerical data, obtained by varying $C_6$. Dotted black line: $0.5\,L/R_b$. Dashed black line: $0.5(L/R_b+1)$. Solid black line: $0.5(L\ind{eff}/R_b+1)$. None of these estimates fits the numerically obtained dependence. (b) Spatial distribution of the Rydberg excitations on the string in excitations per $\mu$m, divided by the total number of excitations.}
 \label{fig:Nexz_Pm}
\end{figure}

Next, we consider the assumption that $N\ind{ryd}\sim m\ind{max}=N/N_b=L/R_b$. In Fig.~\ref{fig:Nexz_Pm}(a), results are shown for the numerically obtained $N\ind{ryd}$ as a function of $C_6$, expressed in units of $L/R_b$ via Eq.~(\ref{eq:Rb}). The dotted line shows $\gamma^*L/R_b$, where $\gamma^*$ is the average fraction to which the super-atoms are excited and was assumed to be 0.5. It can be seen that this estimate does not coincide with the numerically found $N\ind{ryd}$. 
%
%
%

This failure in finite systems is due to two reasons. First, from Fig.~\ref{fig:Nexz_Pm}(b) it can be seen that the distribution of excitations on the string shows strong maxima at both ends. These can be understood in the mean-field picture. The interaction shift of atoms close to the border is smaller than that of atoms in the center since they only have potentially excited neighbors to one side. Consistent with the structure of the pair correlation function, the side maxima have to be followed by minima due to the blockade effect and, towards the center of the string, by further maxima corresponding to the first maximum of $g^{(2)}$. The maxima at the edges of the string lead to a higher excited fraction compared with the case of an infinite string. This effect becomes more dominant as $C_6$ increases since the number of excitations on the string decreases leading to an effectively smaller system. 

Secondly, in a system of length $L$, the assumption of densely packed excitations at distances $R_b$ from each other does not necessarily lead to $N\ind{ryd}=\gamma^*L/R_b$. The number of super-atoms fitting on such a string is rather $L/R_b+1$ due to the excitations sitting at both ends of the string. This would result in $N\ind{ryd}=\gamma^*(L/R_b+1)$ which fits the numerical data much better but overestimates it a bit (dashed black line in Fig.~\ref{fig:Nexz_Pm}(a)). A further obvious effect is connected to the finite density of the gas. The finite atomic line density of $\rho=3\,\mu $m$^{-1}$ implies that it is very unlikely that two atoms are sitting exactly at the trap ends, so the outermost super-atoms are shifted towards the trap center on average by $0.5\,\rho^{-1}$ leading to an efficient trap length of $L\ind{eff}=L-\rho^{-1}$. We therefore show $N\ind{ryd}=\gamma^*(L\ind{eff}/R_b+1)$ (solid black line in Fig.~\ref{fig:Nexz_Pm}(a)) where again $\gamma^*=0.5$ was assumed. This line now reproduces the slope of the numerical data well but still overestimates it. In order to understand this, we should bear in mind that the notion of super-atoms sitting at the edges of the trap is not consistent with the ansatz we used to estimate the blockade radius. $R_b$ is estimated by equating the collective Rabi frequency of $N_b$ atoms to the interaction strength at a distance $R_b$, i.\ e., $\sqrt{N_b}\Omega=C_6/R_b$. A super-atom sitting at the edge of the trap would now contain less atoms, leading to a larger blockade radius. We conclude that the super-atom picture is not adequate to explain the effects of the trap boundaries.   

In general, the validity of universal scaling laws depends on how well length scales are separated from each other. The blockade radius should be much smaller than the trap size but much larger than the intermediate particle distance \cite{weimer2008}. As the simulation in Fig.~\ref{fig:Nexz_Pm} was done at constant atom density and trap size, increasing $L/R_b$ means that the number of atoms per blockade sphere decreases. Thus by increasing $L/R_b$ we effectively move towards the regime of low density, where the interatomic distance is of similar size as the blockade radius. In this regime the predicted scaling laws are not expected to hold.

Fig.~\ref{fig:Nexz_Pm}(a) shows yet another interesting property of small clouds. If $L/R_b$ is smaller than one, the cloud is perfectly blocked and Rabi oscillates between the ground state and the fully symmetric singly excited state as mentioned in Sect.~\ref{sec:dynamics}. Thus the time average of $N\ind{ryd}$ is $0.5$ as confirmed by the simulation data. As $L/R_b$ increases, according to the super-atom picture, the number of excitations should immediately jump to $N\ind{ryd}=1$ as soon as two super-atoms fit into the trap. However, $N\ind{ryd}$ increases smoothly, indicating that  super-atoms cannot be viewed as hard spheres but are rather soft objects. It should also be mentioned that, as the trap size is increased holding density and all other parameters constant, for $L/R_b\gtrsim5.5$, $N\ind{ryd}$ increases linearly with a slope of $0.5/R_b$ confirming the picture of soft super-atoms.

Our above findings explain why the exponent $\nu$ is underestimated if finite-size effects are neglected, and why it is quite difficult in finite systems to find an analytical relation between $N\ind{ryd}$ and $m\ind{max}$ based on simple assumptions. For larger systems, these finite-size effects become negligible relative to the bulk properties of the large ensemble. We thus conclude that the assumptions  $N\ind{ryd}\sim m\ind{max}=N/N_b=L/R_b$ fails for small ensembles, leading to the discrepancy between the numerically obtained scaling parameter $\nu$ and the theoretical prediction in Eq.~(\ref{eq:scaling_exp}).

\begin{figure}[t]
 \centering
 \includegraphics[width=\columnwidth]{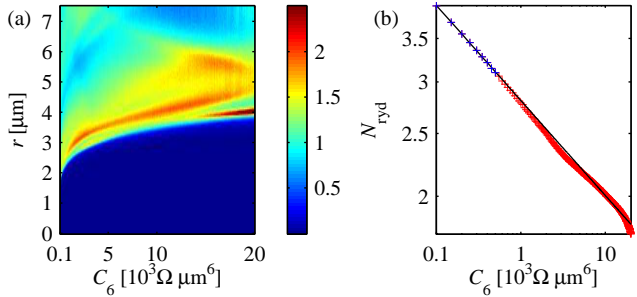}
 \caption{(Color online) Same setup as in Fig.~\ref{fig:g2}, but with periodic boundary conditions. (a) Pair correlation function as a function of $C_6$. Effects of artificial self coupling due to finite system size are visible particularly at high $C_6$. (b) Crosses: Simulated number of Rydberg excitations. Solid line: Fit of algebraic decay. Only the first nine points are used for the fit.}
 \label{fig:pBC}
\end{figure}

\subsection{\label{sec:pBC}Large ensembles via periodic boundary conditions}

If periodic boundary conditions are used, the excitation density is perfectly flat. However, if strong correlations are present, large system sizes are required in order to eliminate artificial self-coupling effects. The latter are also known as aliasing effects that are caused by atoms correlated with a neighboring atom twice, inside the string and across the string boundary. Therefore the string length has to be at least twice as large as the range of the correlations. This is illustrated in Fig.~\ref{fig:pBC}(a). At large values of the interaction strength, for which  the blockade radius is large and correlations are long range, the $g^{(2)}$-function deviates strongly from the case without periodic boundaries. Only at small values of $C_6$ the system seems to become free of artificial self-coupling effects. Note that in contrast to the finite-size effects without periodic boundary conditions discussed above, these self-couplings cannot be observed in a linear 1D ensemble. They could, however, be observed if instead the gas was arranged in a suitable 1D ring trap. To avoid the self-coupling effects, we fitted the algebraic scaling law to the nine points with lowest $C_{6}$ shown in Fig.~\ref{fig:pBC}(b) and obtained $\nu\approx-0.146\pm0.003$. The line density is $\rho=3\,\mu m^{-1}$ as in the fixed-boundary case. For larger densities, i.e., larger $N_b$, the scaling exponent obtained from our numerical simulations approaches the predicted value of $\nu=-2/13$.
This can be understood since at high densities, the nearest-neighbor distances between ground-state atoms is much smaller than $R_b$ while the trap length is much larger than $R_b$. In this case, the system appears homogeneous, as neither the finite trap size nor the coarseness of the atom distribution are relevant.

\subsection{\label{sec:g2}Finite-size effects in the correlation function $g^{(2)}$}

\begin{figure}[t]
 \centering
 \includegraphics[width=\columnwidth]{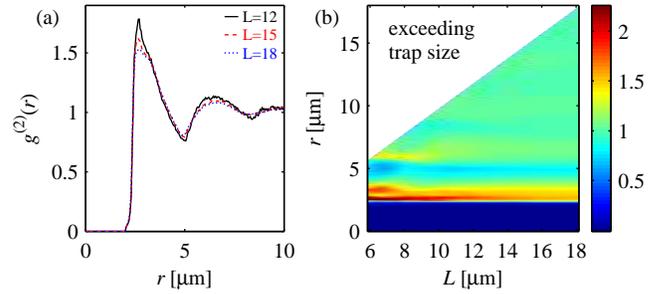}
 \caption{(Color online) Finite-size effects in the $g^{(2)}$-function. If the correlation length is on the order of the trap size, correlations are enhanced by the finite-size effects. Parameters are $C_6=900$\,MHz\,$\mu$m$^6$ and $\rho=3.33\,\mu$m$^{-1}$.}
 \label{fig:g2_trap_size_dep}
\end{figure}

We finally analyze finite-size effects in the pair correlation function. Fig.~\ref{fig:g2_trap_size_dep}(a) shows $g^{(2)}(r)$ for three different trap lengths and a density of $\rho=3.33\,\mu$m$^{-1}$. We notice that the deviation from $g^{(2)}(r)=1$, indicating spatial correlations, decrease slightly with increasing $L$. We now focus on the first maximum of the correlation function, c.\ f.\ Fig.~\ref{fig:g2_trap_size_dep}(a). This maximum is the feature that depends on the trap size most strongly. Looking at this peak more closely, we find a double-peak structure which vanishes in the limit of large very trap size. Splitting up the $g^{(2)}$-function into contributions from the different subspaces with definite excitation number $m$ we find a very pronounced peak in the contribution of the two-fold excited states, which is located at the position of the left peak sub-peak of the first maximum in Fig.~\ref{fig:g2_trap_size_dep}(a). This structure is present for any trap length, but as the relative population of states with a very low number of excitations decreases with increasing trap size, the effect on the structure of the total $g^{(2)}$ becomes weaker for large traps. This explains why the height of the left sub-peak decreases with increasing trap size, whereas the right sub-peak remains constant in size. The structures seen on the $m=2$ subspace can partly be interpreted in the two-atom picture as arising from different excitation channels, as done in Ref.~\cite{gaerttner2012} for the case of $\Delta\gg\Omega$. However, for the $\Omega=0$ case one expects collective effects to be dominant, and thus it is clear that the structure cannot be fully understood from the perspective of the two-atom problem. As expected, correlation peaks are generally enhanced if the system size is smaller than the range of the correlations. Fig.~\ref{fig:g2_trap_size_dep}(b) confirms that $g^{(2)}(r)$ converges to a smooth function in the limit of large $L$. In the simulations with largest system sizes in Fig.~\ref{fig:g2_trap_size_dep}(b), the $g^{(2)}$-function saturates to $1$ over a distance smaller than the trap length. Nevertheless, some features of the correlation function (visible in Fig.~\ref{fig:g2_trap_size_dep}(a) as discussed above) still depend on the trap length.

%

\section{\label{sec:conclusion}Conclusions}

In summary, we have studied the time evolution and the steady-state statistics of the number $N\ind{ryd}$ of Rydberg excited atoms in cylindrical and one-dimensional trap geometries, with and without periodic boundary conditions. 
We predict finite-size effects in the scaling of $N\ind{ryd}$ with the van der Waals coefficient $C_6$ which quantifies the dipole-dipole interaction strength between the excited Rydberg atoms. Our numerical analysis shows that $N\ind{ryd}$ and $C_6$ are still connected by an algebraic scaling, but with a scaling exponent $\nu$ different from the analytically predicted value well-known in the literature.
The analytical prediction relies on two assumptions, an estimate of the blockade radius, and an estimate of the number of excited Rydberg atoms based on the super-atom model. We could show that the estimate of the blockade radius provides a reliable prediction of the position of the first maximum of the pair correlation function obtained from the numerical simulations over a wide range of atom densities as well as interaction strengths. In contrast, the analytical estimate of the number of excited Rydberg atoms did not agree with our numerical results. This was found to be due to the fact that atoms at the ensemble borders can only interact with other atoms to one side, leading to an overall enhanced Rydberg excitation. Additionally, for a continuously varying ensemble size, the number of hard-sphere super-atoms fitting into the volume is not definite, but rather lies between $L/R_b$ and $L/R_b+1$. We also predicted finite-size effects in the pair correlation function $g^{(2)}$, for which we have shown that some of its features  depend on the ensemble size even for lengths $L$ exceeding the distance range over which $g^{(2)}(r)$ is different from one.
In larger ensembles, which we have simulated using periodic boundary conditions, we found that the scaling exponent $\nu$ agrees with the analytical prediction, provided that the density is sufficiently high. Only then, the atomic distribution is sufficiently homogeneous. 
We conclude that two conditions are required to eliminate finite-size effects: First, the condition $R_b\gg \rho^{-1}$ needs to be fulfilled, such that the microscopic structure of the atom distribution becomes irrelevant. Second, $L\gg R_b$ is needed in order to eliminate the effects of the trap boundaries.
We have shown that for trap sizes of $L/R_b<6$ finite size effects significantly modify the dependence of the number of Rydberg excitations on the interaction strength. For typical experimental parameters ($R_b=5\,\mu$m to $10\,\mu$m) this corresponds to $L=30\,\mu$m to $60\,\mu$m which is an experimentally relevant length scale \cite{ditzhuijzen2008, tauschinsky2010, leung2011, viteau2011, kuebler2010}.

\begin{acknowledgments}
This work was supported by University of Heidelberg (Center for Quantum Dynamics, LGFG), by Deutsche Forschungsgemeinschaft (GA 677/7, 8), and  
by the Helmholtz Association (HA216/EMMI).
\end{acknowledgments}

\end{document}